# Confinement and magnetic field effect on chiral ferroelectric nematic in Grandjean-Cano wedge cells

Kamal Thapa[1,2], Olena S. Iadlovska[1,2], Bijaya Basnet[1,3], Hao Wang[1,3], Ayusha Paul[2], James T. Gleeson[2], Oleg D. Lavrentovich[1,2,3*]

[1]*Advanced Materials and Liquid Crystal Institute, Kent State University, Kent, OH 44242, USA.*

[2]*Department of Physics, Kent State University, Kent, OH 44242, USA.*

[3]*Materials Science Graduate Program, Kent State University, Kent, OH 44242, USA*

*Email address: olavrent@kent.edu

**Abstract.** We explore the structure and magnetic field response of edge dislocations in Grandjean-Cano wedge cells filled with chiral mixtures of the ferroelectric nematic mesogen DIO. Upon cooling, the ordering changes from paraelectric in the cholesteric phase $N^*$ to antiferroelectric in the smectic $SmZ_A^*$ and to ferroelectric in the cholesteric $N_F^*$. Dislocations of the Burgers vector $b$ equal the helicoidal pitch $\mathcal{P}$ are stable in all three phases, while dislocations with $b = \mathcal{P}/2$ exist only in the $N^*$ and $SmZ_A^*$. The $b = \mathcal{P}/2$ dislocations split into pairs of $\tau^{-\frac{1}{2}}\lambda^{+\frac{1}{2}}$ disclinations, while the thick dislocations $b = \mathcal{P}$ are pairs of nonsingular $\lambda^{-\frac{1}{2}}\lambda^{+\frac{1}{2}}$ disclinations. The polar order makes the $\tau^{-\frac{1}{2}}$ disclinations unstable in the $N_F^*$ phase, as they should be connected to singular walls in the polarization field. We propose a model of transformation of the composite $\tau^{-\frac{1}{2}}$ line-wall defect into a nonsingular $\lambda^{-\frac{1}{2}}$ disclination, which is paired up with a $\lambda^{+\frac{1}{2}}$ line to form a $b = \mathcal{P}$ dislocation. The $SmZ_A^*$ behavior in the in-plane magnetic field is different from that of the $N_F^*$ and $N^*$: the dislocations show no zigzag instability, and the pitch remains unchanged in the magnetic fields up to 1 T. The behavior is associated with the finite compressibility of smectic layers.

# I. INTRODUCTION

A chiral nematic liquid crystal (N*) forms a helicoidal structure with a periodic twist of the director $\hat{n}$, which describes the average orientation of the rod-like chiral molecules. The director rotates around a helicoidal axis $\hat{\chi}$ remaining perpendicular to it. The distance over which $\hat{n}$ completes a rotation by $2\pi$ is the helical pitch $\mathcal{P}$. Due to the head-tail symmetry, $\hat{n} \equiv -\hat{n}$, the N* periodicity is $\mathcal{P}/2$ [1], Fig.1(a). In confinement, spatial variations of $\hat{n}$ and $\hat{\chi}$ are determined by surface interactions and bulk elasticity. Very often, boundary conditions necessitate the appearance of defects such as dislocations, disclinations, and focal conic domains [2]. When the associated distortions extend over scales much larger than the pitch, the elastic properties of N* are described similarly to those of a smectic A [2].

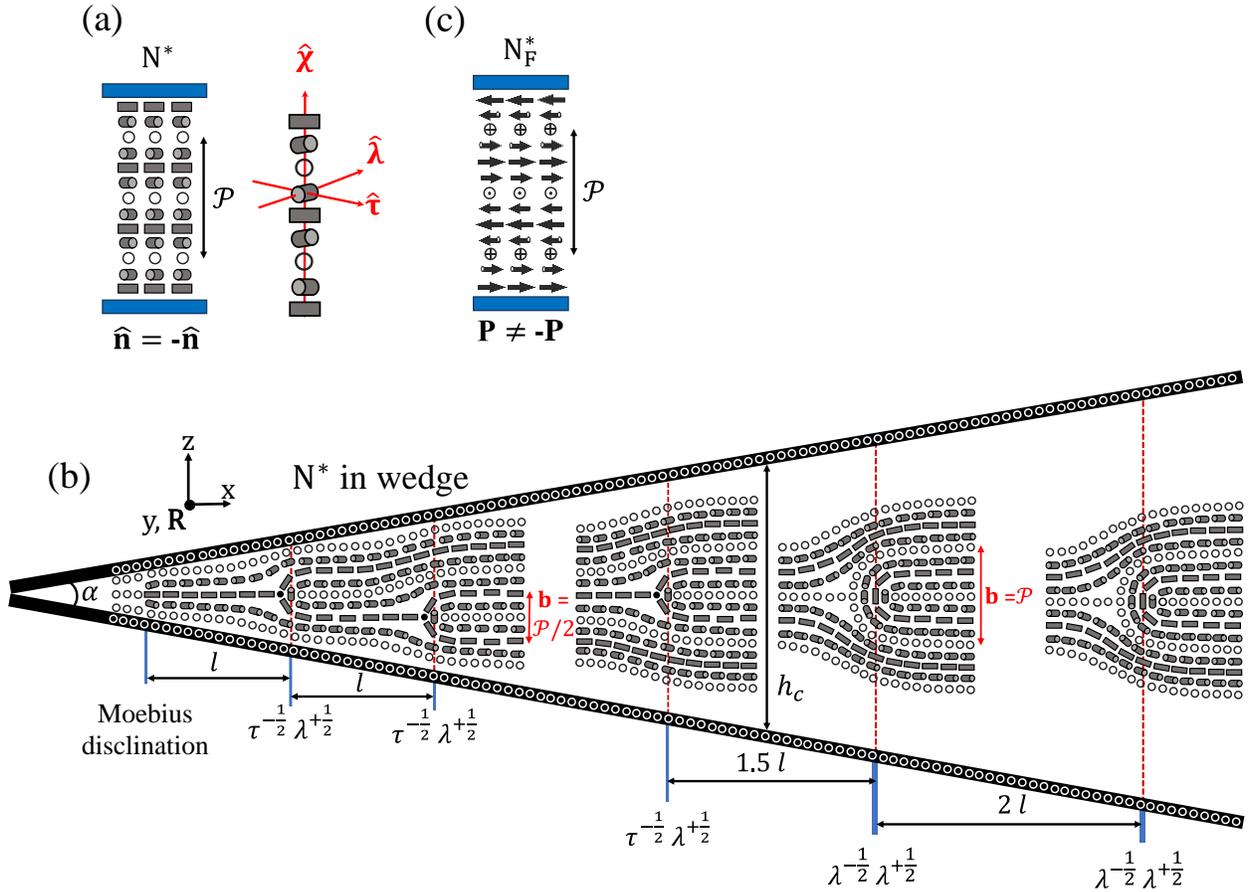

Fig.1. Structural organization of (a) an apolar chiral nematic N* with a helicoidal axis $\hat{\chi}$, local director $\hat{\lambda}$, orthogonal direction $\hat{\tau}$, and period $\mathcal{P}/2$; (b) the N* director field in a Grandjean-Cano wedge with thin $b = \mathcal{P}/2$ and thick $b = \mathcal{P}$ edge dislocations; (c) the chiral ferroelectric nematic $N_F^*$; the structural period is $\mathcal{P}$ rather than $\mathcal{P}/2$.

Confinement-induced $N^*$ structures are often studied in the so-called Grandjean-Cano wedge formed by a pair of glass plates with a small dihedral angle $\alpha$, Fig.1(b). Each glass plate is treated to provide a unidirectional planar alignment. The equilibrium $N^*$ adapts to the varying wedge thickness by introducing line defects [3], Fig.1(b). These lines extend along the direction perpendicular to the thickness gradient and separate neighboring Grandjean zones, defined as the regions within which the number $m$ of director twists by $\pi$ is constant. There are three types of defects [3-16] in the $N^*$ wedge cells.

The line closest to the junction of the plates is a twist disclination, also called a "Moebius disclination," separating a planar untwisted region from the first Grandjean zone in which the director experiences a $\pi$ twist [17]. The second and the third types are "thin" and "thick" dislocations of the Burgers vector $b = \mathcal{P}/2$ and $b = \mathcal{P}$, respectively. Their cores represent pairs of the so-called $\lambda$ and $\tau$ disclinations of strength $\pm 1/2$. The nomenclature has been introduced by Kleman and Friedel [8,9,18] and is based on the triad of axes: $\hat{\boldsymbol{\chi}}$ along the helical axis, $\hat{\boldsymbol{\lambda}}$ along the local director $\hat{\mathbf{n}}$, and $\hat{\boldsymbol{\tau}} = \hat{\boldsymbol{\lambda}} \times \hat{\boldsymbol{\chi}}$, Fig. 1(a); see also a review by Pieranski [16]. In a $\lambda$ disclination of strength $\pm 1/2$, $\hat{\boldsymbol{\tau}}$ and $\hat{\boldsymbol{\chi}}$ rotate by $\pi$ around the core, while $\hat{\boldsymbol{\lambda}}$ is along the core. In a $\tau$ disclination of strength $\pm 1/2$, $\hat{\boldsymbol{\lambda}}$ and $\hat{\boldsymbol{\chi}}$ rotate by $\pi$, while $\hat{\boldsymbol{\tau}}$ is along the core. The $\lambda$ core, extending over a distance comparable to $\mathcal{P}$, is nonsingular and of a lower elastic energy than the singular $\tau$ cores, in which the director is orthogonal to the disclination, Fig.1(b) [8,9,18].

Thin edge dislocations $b = \mathcal{P}/2$ form in the thin part of wedge, Fig.1(b). Their cores split into pairs of $\tau^{-\frac{1}{2}}\lambda^{+\frac{1}{2}}$ disclinations. When the thickness exceeds some critical value $h_c$, the thin dislocations are replaced with thick edge dislocations of the Burgers vector $b = \mathcal{P}$ separating Grandjean zones in which the director twist changes by $2\pi$. The $b = \mathcal{P}$ core splits into a pair of $\lambda^{-\frac{1}{2}}\lambda^{+\frac{1}{2}}$ nonsingular disclinations [14]. The stability of thin vs. thick dislocations and thus the value of $h_c$ are controlled by the balance of orientational and compressional elasticities [14].

Recently, a polar version of the N has been synthesized and characterized [19-24]. This liquid crystal, called a 3D uniaxial ferroelectric nematic ($N_F$) [23], is formed by achiral rod-like molecules with large permanent longitudinal electric dipoles, which align parallel to each other, producing spontaneous macroscopic polarization **P** along $\hat{\mathbf{n}}$. Addition of chiral molecules to the $N_F$ produces a ferroelectric cholesteric phase ($N_F^*$) [25-27], Fig.1(c). The most studied mesogens forming the $N_F$ are abbreviated DIO [20] and RM734 [19]. Because of the polar molecular order, in which **P** and −**P** are not equivalent, the period of $N_F^*$ is the pitch $\mathcal{P}$, which corresponds to a $2\pi$ twist of **P**, Fig.1(c). This principal feature supports $b = \mathcal{P}$ dislocations in the Grandjean-Cano wedge cells, as already observed by Zhao et al. [25], and Nishikawa et al. [26].

The polar nature of $N_F$ manifests itself not only in the bulk structures but also in surface interactions. In conventional N and $N^*$ phases, the surface interactions are usually apolar in the plane of the interface. In contrast, the $N_F$ phase shows polar in-plane ordering: **P** adopts only one equilibrium orientation at a buffed substrate; alignment antiparallel to it is of higher energy [28-30]. Sandwich $N_F$ cells show either a twisted or a monodomain structure depending on whether the planar alignment on the two bounding plates was achieved by buffing along antiparallel directions or parallel directions, respectively [28-30]. The bulk $N_F$ structure in flat samples can also become twisted along the normal to the sample when one surface imposes a unidirectional planar alignment while the other surface allows **P** to orient along any in-plane direction [31]. Such a sample, despite being formed by rod-like molecules with no chemically induced chiral centers, spontaneously splits into domains of alternating left-handed and right-handed twists of **P** in order to reduce the electrostatic energy. This structural chirality caused by $N_F$ electrostatics should be distinguished from the chemically induced chirality in the $N_F^*$ wedges explored in this study.

In addition to the N and $N_F$ phases, DIO exhibits an antiferroelectric smectic-Z phase ($SmZ_A$) with periodic modulation of density and splay of $\hat{\mathbf{n}}$ and **P** [32]. The average $\hat{\mathbf{n}}$, denoted $\overline{\mathbf{n}}$, is parallel to the smectic planes and the sign of **P** alternates from one layer to the next with a period ~18 nm [32]. $SmZ_A$ slabs bounded by two buffed glass plates show two types of alignment: a bookshelf (BK) and parallel alignment (PA), in which the layers are perpendicular and parallel to the plates, respectively, with $\overline{\mathbf{n}}$ along the buffing direction in both cases [32]. These two geometries can be distinguished by their response to an in-plane electric field [32]. The PA readily undergoes a twist Frederiks transition, while the BK resists it [32] since the layers tend to keep equidistance.

The difference in the order parameters, sensitivity to polar surface interactions, and the presence of the $SmZ_A$ phase motivated us to perform a comparative analysis of $N^*$, $SmZ_A^*$, and $N_F^*$ in Grandjean-Cano wedge cells with both parallel and antiparallel assembly of unidirectionally buffed substrates. The direction of buffing is perpendicular to the thickness gradient. The material under study is DIO doped with a chiral additive R1011, which shows the phase sequence Isotropic $-$ N$^*$ $-$ SmZ$_A^*$ $-$ N$_F^*$ upon cooling [29]. The study focuses on five aspects: **(A)** the temperature dependence of the pitch $\mathcal{P}$ measured by the analysis of dislocation networks; **(B)** optical analysis of dislocations' cores; **(C)** zigzag instabilities of dislocations and unwinding of the helicoids by a magnetic field; **(D)** transformation of dislocations in phase transitions; **(E)** alignment of chiral SmZ$_A^*$ phase in confinement. The main findings are as follows:

**(A)** The temperature dependence of $\mathcal{P}$ is weak and non-monotonous in the N$^*$, SmZ$_A^*$ and N$_F^*$ phases, with a minimum in the deep N$^*$ phase. The pitch decreases with the increase of the concentration of chiral dopant.

**(B)** The core width of the $b = \mathcal{P}/2$ dislocations in the N$^*$ and SmZ$_A^*$ is approximately $\mathcal{P}/4$ whereas that of the $b = \mathcal{P}$ dislocations in the N$^*$, SmZ$_A^*$, and N$_F^*$ is approximately $\mathcal{P}/2$. The findings validate the Kleman-Friedel model [8,9] that a $b = \mathcal{P}/2$ core splits into a disclination pair $\tau^{-\frac{1}{2}}\lambda^{+\frac{1}{2}}$ and $b = \mathcal{P}$ splits into $\lambda^{-\frac{1}{2}}\lambda^{+\frac{1}{2}}$.

**(C)** The magnetic field causes zigzag instability of thick dislocations $b = \mathcal{P}$ in both N$^*$ and N$_F^*$. The $b = \mathcal{P}/2$ dislocations remain rectilinear. The pitch of both N$^*$ and N$_F^*$ diverges as the magnetic field increases, in qualitative (but not quantitative) agreement with the theoretical models. In the SmZ$_A^*$, dislocations of any Burgers vector remain straight, in fields up to 1 T.

**(D)** In the N$^*$ and SmZ$_A^*$ phases, thin dislocations $b = \mathcal{P}/2$ occupy the thin part of the wedge, $h < h_c$, while thick dislocations $b = \mathcal{P}$ are located at $h > h_c$. In the N$_F^*$ phase, $\tau^{-\frac{1}{2}}$ disclinations cannot exist as isolated defects and must be attached to a singular wall in the polarization field **P(r)**, which is in contrast to the $\lambda^{\pm\frac{1}{2}}$ disclinations which exist as isolated defects in the N$_F^*$. Restructuring of the $\tau^{-\frac{1}{2}}\lambda^{+\frac{1}{2}}$ core by adding a half-pitch N$_F^*$ layer transforms it into the $\lambda^{-\frac{1}{2}}\lambda^{+\frac{1}{2}}$ core and thus replaces the symmetry-prohibited $b = \mathcal{P}/2$ dislocations with a symmetry-allowed $b = \mathcal{P}$ ones in the N$_F^*$.

**(E)** In the chiral SmZ$_A^*$, the smectic layers are perpendicular to the helicoidal axis $\hat{\chi}$, since this arrangement preserves the layers' equidistance. As a result, in flat sandwich cells, the SmZ$_A^*$ adopts a twisted planar alignment, an analog of the planar alignment geometry of the non-chiral SmZ$_A$.

## II. MATERIALS AND METHODS

We study chiral mixtures abbreviated M0.2, M0.1, and M0.04 containing the ferroelectric material DIO (synthesized as described in ref. [29]) doped with a chiral additive R1011 (Daken Chemicals), in three different weight proportions DIO:R1011 = 99.8:0.2, 99.9:0, and 99.96:0.04, respectively. The mixtures yield different $\mathcal{P}$, according to the approximate dependence $\mathcal{P} \sim 1/c$, where $c$ is the weight concentration of the chiral dopant. The chiral mixture M0.2 is used to study the temperature dependence of $\mathcal{P}$ whereas the magnetic field effects are studied in M0.1. The chiral mixture M0.04 with the largest pitch $\mathcal{P} \approx 60$ µm is used in optical analysis of dislocations cores. The phase sequences upon cooling from the isotropic (I) phase are

I - 166.3°C - $N^*$ - 82.4 °C - $SmZ_A^*$ - 67.2 °C - $N_F^*$ for M0.2,

I - 166.9°C - $N^*$ - 83.4 °C - $SmZ_A^*$ - 68.9 °C - $N_F^*$ for M0.1,

I – 167.0°C - $N^*$ - 83.5 °C - $SmZ_A^*$ - 68.5 °C - $N_F^*$ for M0.04.

The phase diagrams and all other experimental data below are determined with a cooling rate of 0.25°C/min unless specified otherwise.

The wedge cells with a small dihedral angle $\alpha$ are assembled from two glass substrates coated with the polyimide PI2555 (Nissan Chemicals) and unidirectionally buffed as described in Ref. [29]. The buffing directions are perpendicular to the thickness gradient in order to avoid splay which in the $N_F$ causes space charge, and the geometrical anchoring effect, which is a replacement of energetically costly splay with twist [33]. Depending on the parallel or antiparallel orientation of the two buffing directions, the wedge cells are called as being parallel or antiparallel assembled, respectively. The thin part of the wedge is glued by a UV curable glue NOA68 (Norland Products, Inc.) without spacers, while the thickest part contains either 210 µm soda lime glass microspheres (Cospheric LLC.) mixed with NOA68 or a 0.7 mm thick glass plate. The dihedral angle $\alpha$ is measured in empty cells by an interference method [34] using a color filter $(532 \pm 1)$ nm (Thorlabs, Inc.).

The M0.2 birefringence $\Delta n = n_e - n_o$, where $n_e$ and $n_o$ are the extraordinary and ordinary refractive indices, respectively, is measured as the ratio of the optical retardance $\Gamma$ to the cell thickness $h$ in a thin, $h = (1.7 \pm 0.1)$ µm, planar cell with parallel buffing using a PolScope MicroImager (Hinds Instruments) at the wavelength 535 nm. The thickness of the cell is smaller than $\mathcal{P}(T)/4$, so that the N* and N$_F^*$ structures are untwisted [29]. The measurements yield $\Delta n = 0.17$ in the N* (100 °C) and 0.20 in the N$_F^*$ (55 °C).

To determine the temperature dependence of the pitch $\mathcal{P}(T)$ in M0.2, we explore a wedge cell with $\alpha = 1.24°$, exhibiting both $b = \mathcal{P}/2$ and $b = \mathcal{P}$ dislocations [14]. The pitch is calculated as $\mathcal{P}(T) = 2\, l(T) \tan \alpha$, where $l(T)$ is the separation distance between two $b = \mathcal{P}/2$ dislocations, or half the distance between two $b = \mathcal{P}$ dislocations.

To decipher the SmZ$_A^*$ structures, we explore M0.2 in a flat sandwich-like planar cell, $h = (4.1 \pm 0.1)$ μm, such that $\mathcal{P}(T)/4 < h < \mathcal{P}(T)/2$. The twisted planar alignment structure is identified by observing the electrooptical response to an in-plane square-pulse electric field with the frequency 200 Hz [32]. The planar cells are made from two glass substrates; one is plain glass with a buffed PI2555, whereas the other has two ITO electrodes separated by a 1 mm gap, coated with PI 2555 buffed perpendicularly to the edges of the electrodes.

Dislocations in M0.04 are analyzed by optical interference, using a monochromatic filter $(532 \pm 1)$ nm [35]. With $\mathcal{P} \approx 60$ μm, $\Delta n \approx 0.2$, and $\lambda \approx 0.5$ μm, the Mauguin parameter is large, Mau $= \Delta n \mathcal{P}/2\lambda \approx 12$, so that the light polarization can be assumed to follow the rotation of $\hat{\mathbf{n}}$. The incident light with linear polarization at 45° to the rubbing direction **R**, produces both ordinary and extraordinary waves. Their interference, viewed between the crossed polarizers, results in a system of fringes. At the dislocation cores, the molecules are tilted towards the propagating light beam, thus decreasing the effective local birefringence. Perturbations of fringes at the locations of dislocations allow us to characterize their cores.

The temperature is controlled by a hot state HCS302 and a controller mK2000 (both Instec, Inc.) with an accuracy of 0.01 °C. The textures are taken using a polarizing optical microscope Nikon OPTIPHOT2-POL (Nikon Inc.) equipped with a QImaging camera. The magnetic experiments are performed using M0.1 wedge cells. A larger $\mathcal{P}$ of this mixture allows one to explore the helicoid unwinding by relatively low fields. In the magnetic field experiments, the temperature is controlled within $\pm 0.1°C$ by a custom-made controller and a hotplate.

In all experiments, the chiral mixtures are filled into the wedge and flat cells in the isotropic phase, by capillary action. All measurements are taken below 120 °C in order to avoid degradation of DIO [20,36].

## III. RESULTS AND DISCUSSION

### A. Temperature dependence of the pitch in M0.2

*Temperature dependence of the pitch upon cooling.* A Grandjean-Cano wedge of a known dihedral angle $\alpha$ allows one to measure the cholesteric pitch $\mathcal{P} = 2\,l \tan\alpha$ by measuring the distance $l$ separating thin $b = \mathcal{P}/2$ dislocations and $2l$ separations between thick $b = \mathcal{P}$ dislocations. We use a parallel assembly with $\alpha = 1.24\,°$. The thin dislocations are stable in the thin part of the sample, $h < h_c$, and the thick ones at $h > h_c$, where $h_c \approx 21$ µm is measured at the mid-way distance between the last $b = \mathcal{P}/2$ dislocation and the first $b = \mathcal{P}$ dislocation, Fig.2(a). Both dislocation types preserve their shape during the N* - SmZ$_A^*$ transition and in the SmZ$_A^*$ phase. One can also observe dislocations with $b = 0$ oriented along the thickness gradient, Fig.2(a). The N$_F^*$ phase exhibits only thick dislocations $b = \mathcal{P}$, Fig.2(a).

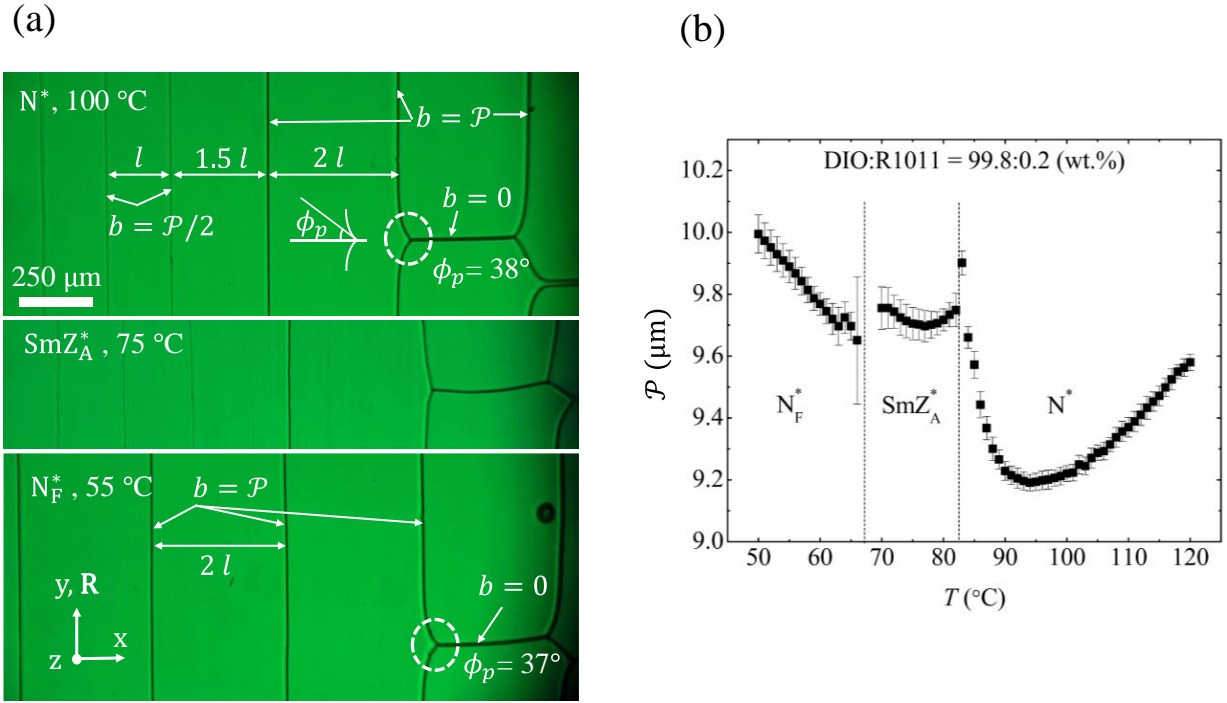

Fig. 2. Measurements of the pitch: (a) lattice of edge dislocations in a parallel buffed wedge cell ($\alpha = 1.24\,°$) in the N*, SmZ$_A^*$, and N$_F^*$ phases; (b) temperature dependence of the helical pitch in M0.2.

Fig. 2(b) shows the temperature dependence $\mathcal{P}(T)$, which is weakly nonmonotonous, without significant divergence at the N*- SmZ$_A^*$ transition. This behavior is different from the divergence of $\mathcal{P}(T)$ near the N*- SmA phase transition [37-39], in which case the twist of $\hat{\mathbf{n}}$ is incompatible with the layered structure. The finite $\mathcal{P}(T)$ and the preserved structure of edge dislocations indicate that twists are allowed in the SmZ$_A^*$ since $\hat{\mathbf{n}}$ and **P**, being in the plane of smectic layers, rotate from one layer to the next.

### B. Dislocation cores in M0.04

When a wedge cell is illuminated with a 532 nm monochromatic light and viewed between crossed polarizers, with the polarizer at 45° to the rubbing direction **R**, one observes birefringence fringes of equal width in the regions between edge dislocations; these are perturbed by the dislocation cores, Fig. 3 [35].

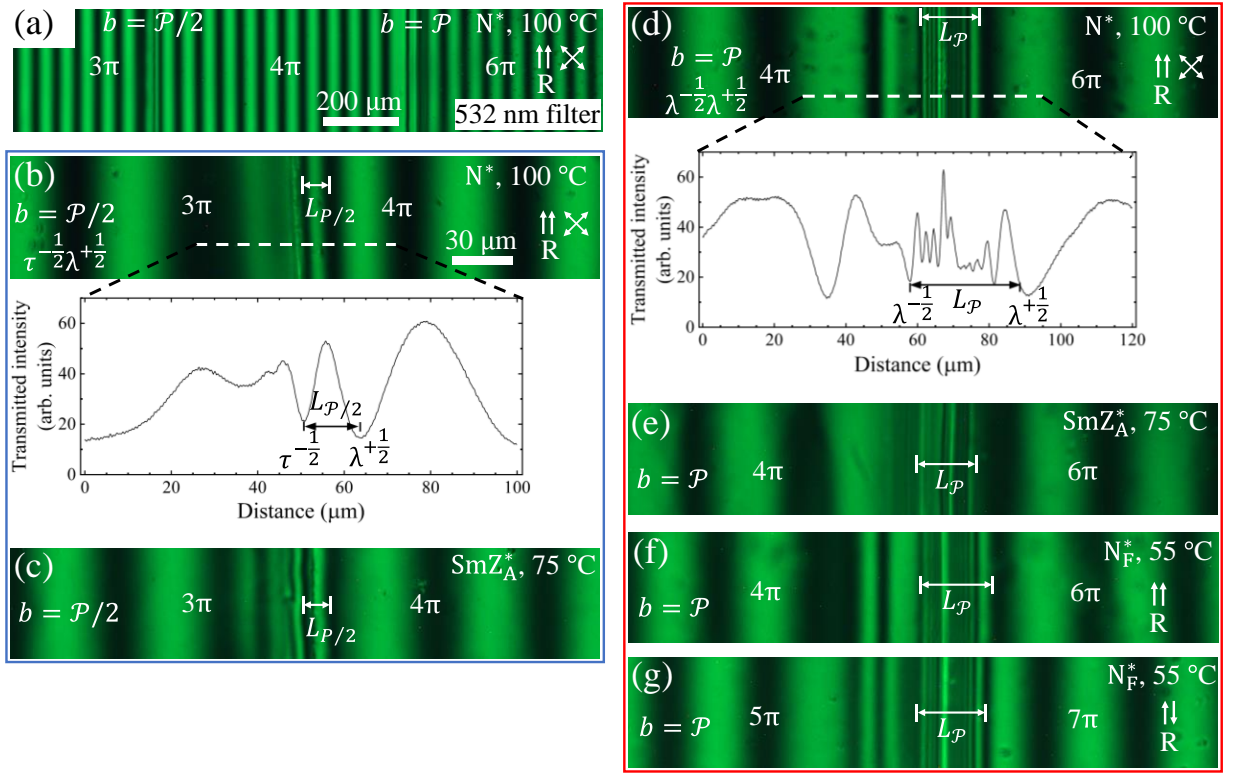

Fig. 3. Birefringence fringes in M0.04 wedge cells with parallel (a-f) and antiparallel assembly (g): (a) thin $b = \mathcal{P}/2$ and thick $b = \mathcal{P}$ dislocations in the N*; (b) a high-magnification image of a thin $b = \mathcal{P}/2$ core in the N*; the inset shows the transmitted light intensity along the white dotted line; (c) the same $b = \mathcal{P}/2$ core in the SmZ$_A^*$; (d) a high-magnification image of a thick $b = \mathcal{P}$ core in the N*; the inset shows the transmitted intensity along the white dotted line; (e) the same $b = \mathcal{P}$

core in the SmZ$_A^*$ and (f) in the N$_F^*$; (g) $b = \mathcal{P}$ in the N$_F^*$ wedge with antiparallel assembly. Dihedral angle $\alpha = 3.39\,°$ in (a-f) and $\alpha = 3.36\,°$ in (g).

The equidistant birefringence fringes in the N* are perturbed by the dislocations, Fig.3(a). At the singular core of $\tau^{-1/2}$ disclination, Fig.3(b), the effective birefringence is discontinuous, which results in light diffraction and an image of the core as a sharp black thread, similar to the $\tau^{-1/2}$ textures described in other N* material by Malet and Martin [35]. As one moves towards the $\lambda^{+1/2}$ disclination, the tilt of the helicoidal axis, and effective birefringence change continuously, producing a broader intensity minimum. The distance $\mathcal{P}/4$ between the two intensity minima is associated with the core extension $L_{\mathcal{P}/2}$ of the $b = \mathcal{P}/2$ dislocation, i.e., the distance between $\tau^{-1/2}$ and $\lambda^{+1/2}$ disclinations in N*, Fig. 3(b), and in the SmZ$_A^*$, Fig. 3(c).

If the thick $b = \mathcal{P}$ dislocation core splits into a pair of nonsingular $\lambda^{-1/2}$ and $\lambda^{+1/2}$ disclinations, as predicted [8,9], then the birefringence should change continuously and rapidly because of the continuous reorientation of the helicoidal axis [35]. This indeed is observed in the interference patterns with multiple maxima and minima, Fig.3(d-g). The extension $L_\mathcal{P}$ of this strongly perturbed zone is close to $\mathcal{P}/2$ in the N*, Fig.3(d), SmZ$_A^*$, Fig. 3(e), and N$_F^*$, Fig. 3(f,g). Table (I) shows the pitch $\mathcal{P}$, measured as $\mathcal{P} = 2\,l \tan \alpha$, and core extensions of both types of dislocations in all three chiral phases. We conclude that the dislocation cores split into the disclination pairs as predicted by Kleman and Friedel [8,9]: $b = \mathcal{P}/2$ splits into a disclination pair with a singular $\tau^{-\frac{1}{2}}$ and nonsingular $\lambda^{+\frac{1}{2}}$, while $b = \mathcal{P}$ splits into a nonsingular pair $\lambda^{-\frac{1}{2}}\lambda^{+\frac{1}{2}}$.

Table (I). Pitch $\mathcal{P}$ and extensions $L_{\mathcal{P}/2}$ of $b = \mathcal{P}/2$ and $L_\mathcal{P}$ of $b = \mathcal{P}$ dislocation cores in M0.04.

| Phase | $\mathcal{P}$ (μm) | $\mathcal{P}/4$ (μm) | $\mathcal{P}/2$ (μm) | $L_{\mathcal{P}/2}$ (μm) | $L_\mathcal{P}$ (μm) |
|---|---|---|---|---|---|
| N*, 100 °C | 58±1 | 14.5 | 29 | 14 | 30 |
| SmZ$_A^*$, 75 °C | 60±1 | 15 | 30 | 14 | 31 |
| N$_F^*$, 55 °C | 64±2 | | 32 | | 35(parallel) |
| | | | | | 33(antiparallel) |

## C. Magnetic field effects in M0.1

Thin and thick edge dislocations in a conventional N* Grandjean-Cano wedge show dramatically different responses to the magnetic field **B** applied along the thickness gradient ($x$-axis) [7]: $b = \mathcal{P}$ lines experience a zig-zag instability, while $b = \mathcal{P}/2$ ones remain rectilinear. The zigzag instability occurs when the field exceeds a critical value $B_{zz} \approx \frac{B_c}{2}$, where

$$B_c = \frac{\pi^2}{\mathcal{P}} \sqrt{\frac{\mu_0 K_2}{\chi_a}} \qquad (1)$$

is the critical field unwinding the N* helicoid [40,41], $K_2$ is the twist elastic constant, $\mu_0$ is the vacuum magnetic permeability, and $\chi_a > 0$ is the anisotropy of diamagnetic susceptibility. The DIO mixture M0.1 shows a similar behavior in the N* phase, Fig.4(a), but presents a plethora of new effects in the two other chiral phases, SmZ$_A^*$, Fig.4(b) and N$_F^*$, Figs.4(c-e).

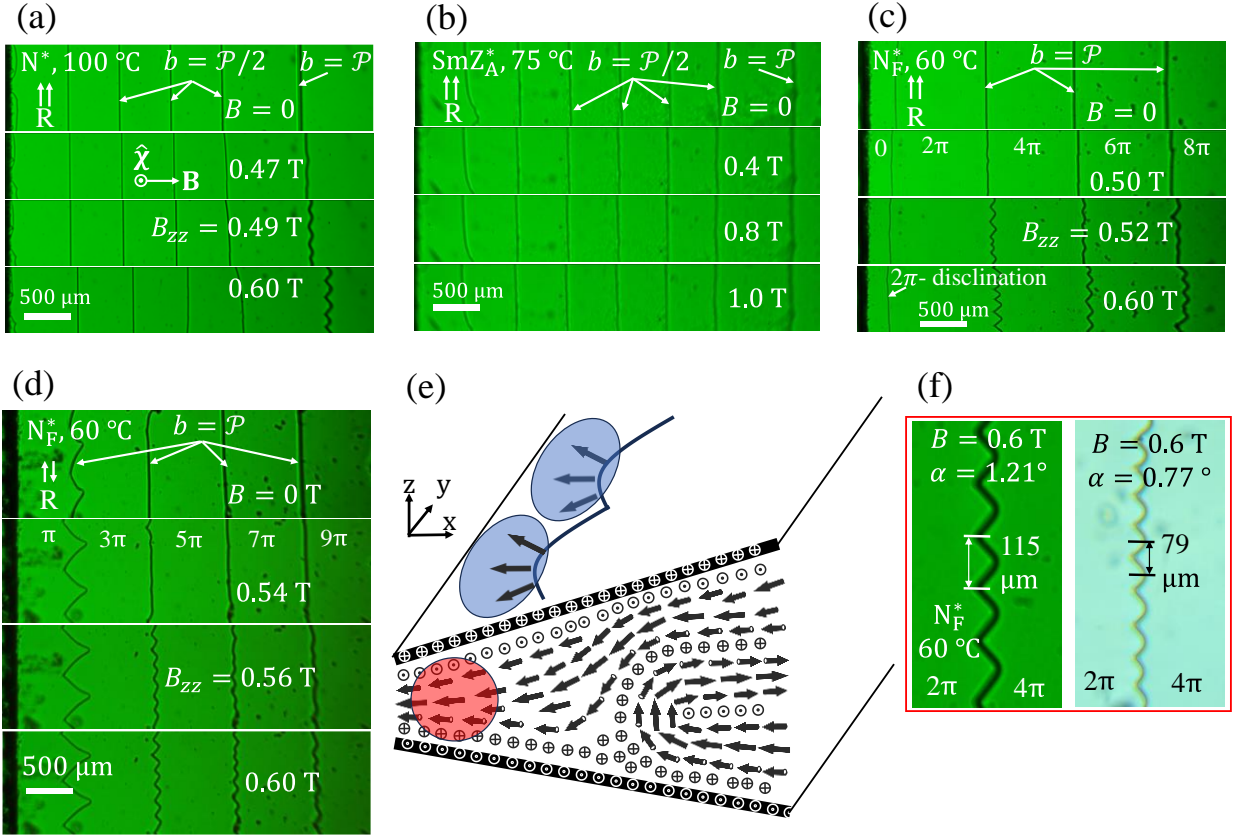

Fig.4. Dislocations in the magnetic field **B**: (a) parallel assembly N* wedge ($\alpha = 1.51°$), $b = \mathcal{P}/2$ dislocations remain straight, $b = \mathcal{P}$ dislocations adopt zigzag shapes above $B_{zz}$; (b) the same

wedge, all dislocations in the SmZ$_A^*$ remain straight; (c) the same wedge, all $b = \mathcal{P}$ dislocations in the N$_F^*$ change to zigzag shapes above $B_{zz}$; the $2\pi$-disclination near the edge of the wedge remains straight; (d) antiparallel assembly wedge ($\alpha = 1.79°$) in the N$_F^*$; $b = \mathcal{P}$ dislocations change to zigzag shapes above $B_{zz}$; note the arched shape of the first disclination in the absence of the field, which is attributed to the splay-cancellation mechanism illustrated in (e); see text for details; (f) period of zigzag undulations in parallel rubbed N$_F^*$ wedge cells increases with the dihedral angle $\alpha$; $B = 0.6$ T. Mixture M0.1.

***Zigzag instability in parallel assembly wedge, Fig.4(a,c,f).*** The magnetic field up to $\approx 1$ T has no effect on the shape of dislocations $b = \mathcal{P}/2$ in the N$^*$ (except for the shift associated with the pitch increase); these remain straight. The $b = \mathcal{P}$ dislocations adopt a zigzag shape in both the N$^*$, Fig. 4(a), and N$_F^*$, Fig. 4(c), when the field increases above $B_{zz}$, equal 0.49 T in the N$^*$ at 100 °C and $\approx 0.54$ T in the N$_F^*$ at 60 °C. The period of zigzag wave increases with the dihedral angle $\alpha$, Fig. 4(f). In the N$_F^*$ phase, the $2\pi$-disclination that separates the untwisted region from the $2\pi$-twist Grandjean zone, Fig.4(c), remains straight; the effect is analyzed in section D. In a striking difference with the two other phases, the $b = \mathcal{P}/2$ and $b = \mathcal{P}$ dislocations in the SmZ$_A^*$ show no zigzag instabilities in the field up to 1 T, Fig.4(b).

The shapes of N$^*$ dislocations in the magnetic field, Fig. 4(a), were explained by Kleman [42] as a result of the core splitting. As established above, a $b = \mathcal{P}$ core splits into a nonsingular pair $\lambda^{-\frac{1}{2}}\lambda^{+\frac{1}{2}}$ and a $b = \mathcal{P}/2$ core splits into a pair $\tau^{-\frac{1}{2}}\lambda^{+\frac{1}{2}}$. The splitting is related to the fact that the product of two opposite rotations by $\pi$ and $-\pi$ along two parallel axis is a translation [8,9]. In the space between the two disclinations, the helicoidal axis $\hat{\chi}$ is along the thickness gradient and is parallel to the applied field **B**, Figs.1(b), 4(a,b). Since $\chi_a > 0$, this orientation maximizes the energy of diamagnetic coupling. The energy is reduced if $\hat{\chi}$ tilts away from **B**, thus forming a zigzag. The critical field causing a zigzag instability is [42] $B_{zz} \approx B_c/2$ for the $\lambda^{-\frac{1}{2}}\lambda^{+\frac{1}{2}}$ pair and $B_{zz} \approx B_c$ for the $\tau^{-\frac{1}{2}}\lambda^{+\frac{1}{2}}$ pair. The larger value of $B_{zz}$ in the case of a $\tau^{-\frac{1}{2}}\lambda^{+\frac{1}{2}}$ pair is explained by a shorter separation between the disclinations and by the singular character of the $\tau^{-\frac{1}{2}}$ core, which results in its high line tension $E_{c,\tau\lambda} \approx \frac{\pi}{2} K ln \frac{\mathcal{P}}{4r_c}$, where $K$ is an average value of the Frank elastic constants and $r_c$ is the core radius of molecular dimension. The core of $\lambda^{-\frac{1}{2}}\lambda^{+\frac{1}{2}}$ pairs is non-singular, with a much lower elastic energy per unit length, on the order of $\sim K$ [42].

As shown in the next section and in Fig.5, the magnetic field that unwinds the cholesteric helix is $B_c = 0.96$ T in the N$^*$ and 1.05 T in the N$_F^*$. In both phases, $B_{zz}/B_c \approx 0.5$, remarkably close to $1/2$ predicted by Kleman [42].

As indicated above, the SmZ$_A^*$ dislocations remain rectilinear in the field, Fig.4(b), which is explained by finite compressibility of the SmZ$_A^*$ layers. The equidistance of SmZ$_A^*$ layers is violated near the $-1/2$ disclination cores, where $\hat{\mathbf{n}}$ experiences alternating bend and splay, Fig.1(b). In the SmZ$_A^*$, splay of $\hat{\mathbf{n}}$, which is parallel to the smectic layers, means that the layers cannot keep their thickness constant. The situation is opposite in a conventional smectic A, in which $\hat{\mathbf{n}}$ is normal to the smectic layers and it is bend and twist of $\hat{\mathbf{n}}$ that are prohibited by the layered structure [2]. In contrast to the $-1/2$ cores, the $+1/2$ disclination cores in the SmZ$_A^*$ can be constructed exclusively with the bend of $\hat{\mathbf{n}}$, which supplements the intrinsic twist, Fig.1(b), thus preserving the equidistance. A zigzag shape of dislocations in the SmZ$_A^*$ would imply additional splay in the $xy$ plane of the sample, which is not compatible with the requirement of layers' equidistance. Smectic layering is the reason why the SmZ$_A^*$ dislocations of any Burgers vector preserve their rectilinear shape in the magnetic field.

***Zigzag instability in antiparallel assembly wedge, Fig.4(d,e).*** The first dislocation in the thinnest part of the antiparallel assembly wedge, which separates the zones with π- and 3π-twists, adopts a peculiar wavy shape even at $B = 0$. The apparent reason is the splay of polarization **P**, necessitated by the wedge geometry and by the fact that in the bisecting plane, **P** is collinear with the thickness gradient, Fig.4(e). N$_F$ and N$_F^*$ textures tend to avoid splay since it produces a bound electric charge of a density $\rho_b = -\text{div } \mathbf{P}$ and increases the electrostatic energy [43]. The bound charge can be reduced by the "splay cancellation" mechanism described in the context of the director for N droplets by Press and Arrott [44] and for hybrid aligned [45] and suspended [46] N films. The space charge produced by splay in the vertical plane $xz$ can be reduced by additional splay of an opposite polarity in the $xy$ plane, Fig.4(e). Imagine, following Ref. [45], that **P** experiences only splay, forming straight lines perpendicular to a set of curved surfaces Σs. Each point at a surface Σ is characterized by two principal radii of curvature $R_1$ and $R_2$ that define the mean curvature $\frac{1}{R_1} + \frac{1}{R_2}$ and the Gaussian curvature $\frac{1}{R_1 R_2}$. The divergence of polarization is related to the mean curvature, div $\mathbf{P} \propto \frac{1}{R_1} + \frac{1}{R_2}$ [2]. The signs of $R_1$ and $R_2$ depend on the orientation of vectors $\mathbf{R}_1$ and $\mathbf{R}_2$ with respect to the chosen normal to $\Sigma$. For a spherical $\Sigma$, the radii are of the same sign, $R_1 R_2 > 0$, while they are of opposite signs, $R_1 R_2 < 0$, for saddle-like surfaces such as a hyperbolic paraboloid. In the discussed geometry, Fig.4(e), let $\mathbf{R}_1$ be the radius of curvature induced by the wedge in the $xz$ plane, producing an electric charge density $\rho_b \propto \left|\frac{1}{R_1}\right|$. If there is additional splay with $\mathbf{R}_2$ in the $xy$ plane such that $R_1 R_2 < 0$, this space charge is reduced, $\rho_b \propto \left|\frac{1}{R_1}\right| - \left|\frac{1}{R_2}\right|$. In Fig. 4(e), a red disk marks a cloud of positive space charges and blue disks mark the negative charges that compensate for the positive charges. Therefore, the wavy shape of the first disclination can be caused by the wedge-induced splay of polarization in the $xz$ plane and a reduction of the associated space charge by additional splay in the $xy$ plane. The effect should be most pronounced in the Grandjean zone with one $\pi$-twist, as in that zone, the polarity of splay is the same everywhere. Grandjean zones with a higher number of twists would show splay of alternating signs as one moves along the $z$-axis, which diminishes the need for additional splay. The negative sign of the vertical $xz$ splay, $\frac{\partial P_z}{\partial z} < 0$, in Fig. 4e is accompanied by the splay-cancelling positive $xy$ splay $\frac{\partial P_y}{\partial y} > 0$ within an arch that bulges towards the thin part of wedge. The arches are separated by cusp-like regions of high curvature in which $\frac{\partial P_y}{\partial y}$ and $\frac{\partial P_z}{\partial z}$ are both negative. The space charge in these regions is most likely screened by freely moving ions always present in liquid crystals. The advantage of the combined geometrical and ionic screening of the bound charge $\rho_b = -\text{div } \mathbf{P}$ is that the average bulk concentration of ions can be low but still sufficient to accumulate in a limited portion of space to screen $\rho_b$ there. The tendency of the first dislocation in Fig.4(d,e) to lower its length to minimize the core energy also contributes to the asymmetric arched shape. An increased angular extension of a splay-cancelling arch implies a longer total length and thus should be limited. The interplay of splay cancellation and ionic screening requires further studies.

*__Unwinding the helix.__* The magnetic field dependence of the pitch $\mathcal{P}(B)$ in the N* (100 °C), SmZ$_A^*$ (75 °C), and N$_F^*$ (60 °C) phases of M0.1 are determined with the field increasing and decreasing in small 0.04 T increments, with 8 hours equilibration at each step, Fig.5. The equilibration results in a practically hysteresis-free behavior. The pitch increases with $B$ in both the N* and $N_F^*$, but remains practically constant in the explored range in the $SmZ_A^*$.

The critical field at which the helix unwinds completely is determined as the field at which the last dislocation is expelled from the sample: $B_c = 0.96$ T in the N* phase and 1.05 T in the $N_F^*$ phase. Using the measured $B_c$ and the off-field pitch $\mathcal{P} = 23.8$ μm in the N* and 25.6 μm in the $N_F^*$ phases of M0.1, one estimates $K_2/\chi_a = 1.1 \times 10^{-6}$ SI for N* and $1.5 \times 10^{-6}$ SI for $N_F^*$. Since $\chi_a \approx 10^{-6}$ in most liquid crystals, $K_2 \approx 1$ pN in the N* and $\approx 1.5$ pN in the $N_F^*$.

In contrast to the helix unwinding in both the N* and $N_F^*$, the pitch of $SmZ_A^*$ does not increase in the field up to 1 T, Fig.5. The apparent reason is the smectic layering of $SmZ_A^*$, which prevents dislocations from experiencing zigzag instability, as discussed above, Fig.4(b). Apparently, the dislocations form a barrier for the realignment in the Grandjean zones so that the pitch does not diverge. It would be of interest to explore $SmZ_A^*$ samples in the fields above 1 T.

The unwinding of N* by a magnetic field is well known since the early theoretical works by de Gennes [40] and Meyer [41] and experiments by the number of research groups [47,48]. The theory considers an infinite sample with a helicoidal twist of the director, $n_x = \cos\varphi(z), n_y = \sin\varphi(z), n_z = 0$, and the free energy comprised of the elastic twist term and the diamagnetic coupling to the applied magnetic field $\mathbf{B} = \{B, 0, 0\}$:

$$F = \int_{-\infty}^{\infty} \left[\frac{1}{2}K_2\left(\frac{d\varphi}{dz} - q_0\right)^2 - \frac{\chi_a}{2\mu_0}B^2\sin^2\varphi\right]dz, \qquad (2)$$

where $q_0 = 2\pi/\mathcal{P}_0$, $\mathcal{P}_0$ is the cholesteric pitch in the absence of the field, and $\varphi$ is the angle between the director and **B**. This functional is applicable to both the N* and $N_F^*$ phases. The analysis predicts the value of the critical field $B_c$ of helix unwinding, Eq.(1), and the pitch dependence $\mathcal{P}(B)$ on the applied magnetic field [40],

$$\frac{\mathcal{P}(B)}{\mathcal{P}_0} = \frac{4}{\pi^2}\mathcal{K}(k^2)\mathcal{E}(k^2), \qquad (3)$$

where $\mathcal{K}(k^2) = \int_0^{\pi/2} \frac{d\psi}{\sqrt{1-k^2\sin^2\psi}}$ and $\mathcal{E}(k^2) = \int_0^{\pi/2} \sqrt{1-k^2\sin^2\psi}\,d\psi$ are the complete elliptic integrals of the first and the second kind, respectively. Note that Eq.(3) contains no adjustable parameters; it predicts that the pitch diverges with $\mathcal{K}(k^2)$ as the field approaches $B_c$, when $k^2 = 1$ and $\mathcal{E}(k^2) = 1$. In weak fields, $\frac{\mathcal{P}(B)}{\mathcal{P}_0} \approx 1 + \frac{\chi_a^2 \mathcal{P}_0^4 B^4}{2^9 \pi^4 \mu_0^2 K_2^2}$, and near $B_c$, the pitch diverges as $\frac{\mathcal{P}(B)}{\mathcal{P}_0} = \frac{4}{\pi^2} \ln \frac{4B_c}{\sqrt{B_c^2 - B^2}}$ [49].

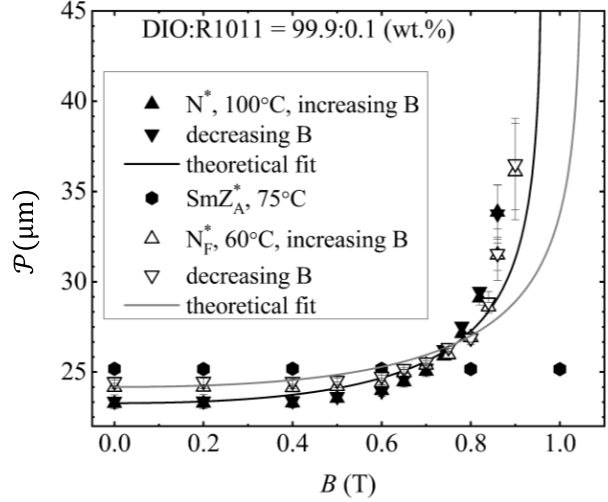

Fig. 5. Magnetic field dependence of helical pitch in parallel buffed wedge cell ($\alpha = 1.51°$) in the $N^*$, $SmZ_A^*$, and $N_F^*$ phases of M0.1.

The theory has been confirmed by experiments, in which the sample represented a fingerprint texture [48] or the Grandjean-Cano wedge cell [47]. Our experimental data, Fig.5, are in qualitative agreement with the model, Eq.(3), for both the $N^*$ and $N_F^*$. The pitch $\mathcal{P}$ in the intermediate fields, from 0.4 T to 0.7 T, is somewhat below the theoretical prediction; a similar discrepancy has been observed in the electric field experiments [50]. At high fields, above 0.8 T, $\mathcal{P}$ is larger than the theory prediction, Fig.5. The absence of quantitative agreement is likely caused by the confinement effects. One of these is surface anchoring, which makes the pitch change a step-like process [51-55]. Another feature is the variations of the structure along the $x$ and $z$ axes, with periodic compressions and dilations and the presence of extended defect cores in which the $\hat{\chi}$ axis is parallel rather than perpendicular to the applied field. These factors contribute to the discrepancies between the experiment and the theory, especially in the relatively thin samples explored in our work. The important role of the dislocation presence in shaping the $\mathcal{P}(B)$ function is further supported by the data on $SmZ_A^*$ in Fig.5. If the $SmZ_A^*$ were an infinite sample without dislocations, then Eqs. (2) and (3) would describe its unwinding since the only deformation is twist of the director from one smectic layer to the next, which does not change the thickness of the layers. However, when the smectic layers are perpendicular to the field, the unwinding is hindered, and the twist-only model is not applicable.

### D. Structural features of chiral phases in wedge cell of M0.2

A detailed exploration of Grandjean-Cano $N^*$ structures was achieved by the fluorescent confocal polarizing microscopy (FCPM) [14]. FCPM requires materials with a low birefringence, ~0.1 or less. High birefringence of DIO diminishes the resolving power of FCPM, which explains why our study is limited by a conventional optical microscopy. The results above suggest the Grandjean-Cano wedge DIO $N^*$ structure as shown in Fig.1(b) and the $N_F^*$ structure as schematized in Fig. 6(a) for parallel assembly and Fig.6(b) for antiparallel assembly. As compared to the $N^*$ wedge in Fig.1(b), there are no $b = \mathcal{P}/2$ dislocations in the $N_F^*$, which are prohibited since the states **P** and $-$**P** are not equivalent to each other. Below we describe the mutual transformations of the $N^*$, $SmZ_A^*$, and $N_F^*$ dislocation lattices upon cooling and heating.

*<u>Dislocations upon cooling; parallel assembly.</u>* The Grandjean-Cano $N^*$ textures of DIO mixtures are similar to the ones described for conventional $N^*$ materials [7,14,17,35]. Namely, the thinnest part of the wedge exhibits an untwisted $N^*$ with $\hat{\mathbf{n}}$ along the rubbing direction [14], Fig.1(b) and 7(a). As the thickness increases, but remains below a critical value $h_c$ [14], one observes thin edge dislocations $b = \mathcal{P}/2$ separating Grandjean zones in which the director twist differs by $\pi$, Fig.7(a). During the $N^*$- $SmZ_A^*$ transition and in the $SmZ_A^*$, these dislocations do not change, except for the increase of the separation $l = \mathcal{P}/(2 \tan \alpha)$, caused by the increase of $\mathcal{P}$, Fig.2(b). The reason is that the $SmZ_A^*$ is antiferroelectric, thus the disclination cores can still afford $\pi$ rotations.

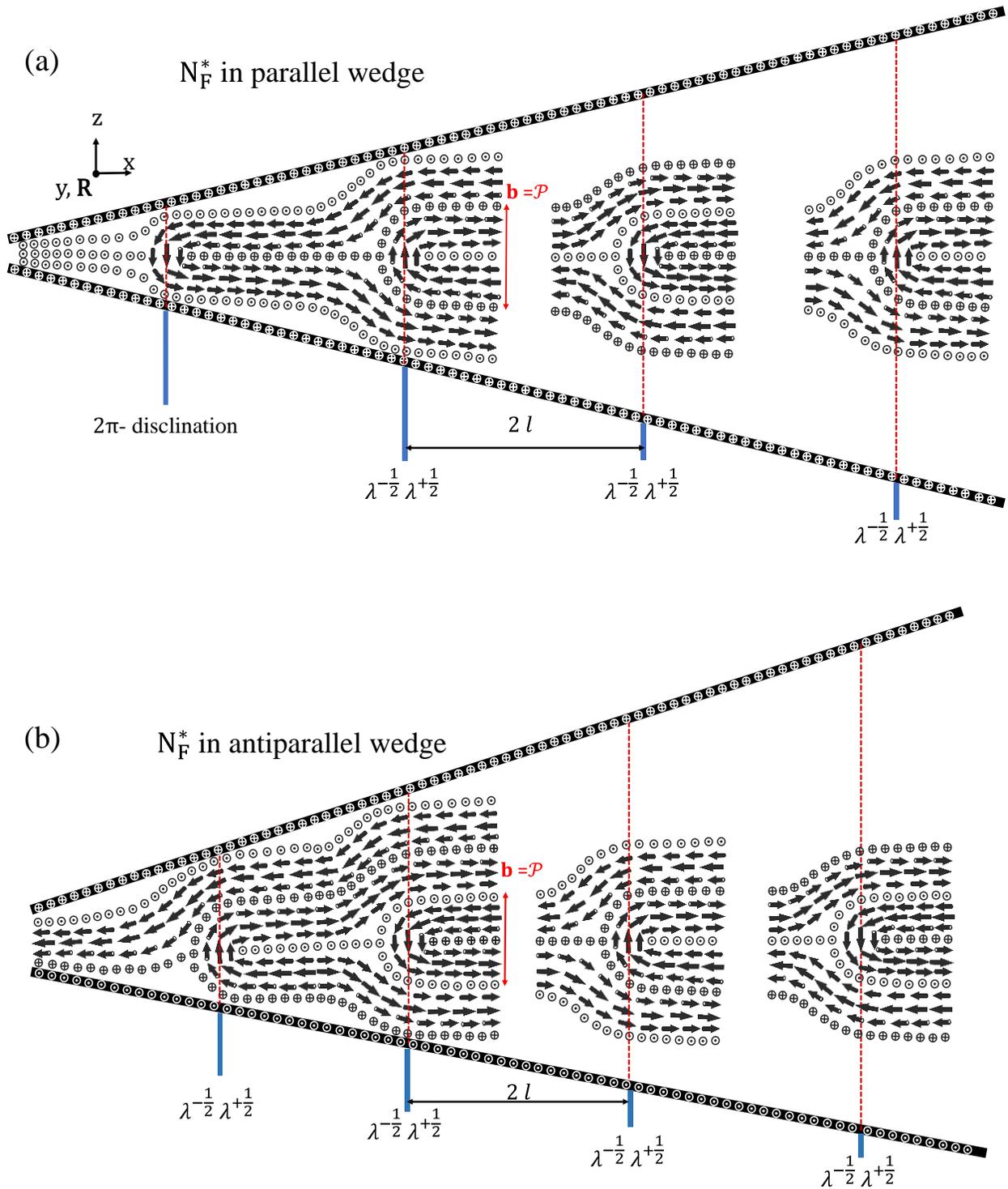

Fig.6. Structural schemes of the Grandjean-Cano wedge with (a) parallel assembly of buffed polyimide alignment layers and (b) antiparallel assembly. The dislocations represent split pairs of nonsingular $\lambda^{-\frac{1}{2}} \lambda^{+\frac{1}{2}}$ disclinations.

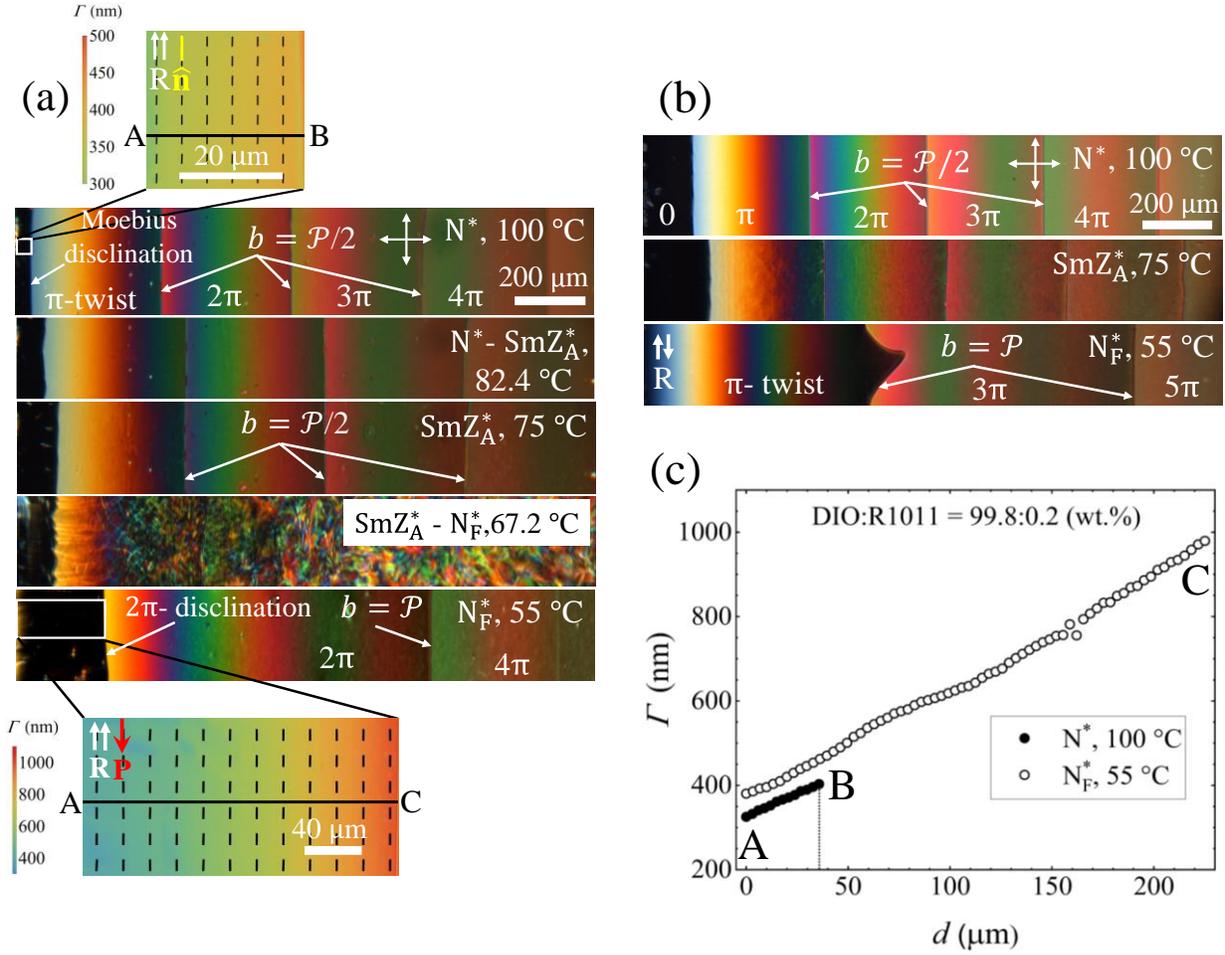

Fig.7. Transformations of thin edge dislocations $b = \mathcal{P}/2$ in the thin part of the wedge on cooling in (a) parallel assembly, $\alpha = 0.74°$ and (b) antiparallel assembly $\alpha = 0.77°$. The edge dislocations preserve their structure upon the $N^*$ to $SmZ_A^*$ transition but increase the separation distances since the pitch increases. In the $N_F^*$, the number of dislocations is reduced by a factor of 2 and all of them are of the Burgers vector $b = \mathcal{P}$; (c) optical retardance of the untwisted regions (a) grows linearly with distance $d$ along the thickness gradient in the $N^*$ (100 °C), line AB, and the $N_F^*$ phase (55 °C), line AC. Mixture M0.2.

Once the temperature is reduced below the $SmZ_A^*$ - $N_F^*$ transition point, the Moebius $\pi$-disclination, and the first edge dislocation $b = \mathcal{P}/2$ merge to form a $b = \mathcal{P}$ dislocation; the untwisted zone moves towards the thicker part of the wedge, Fig 7(a). Insets in Fig.7(a) map the retardance $\Gamma$ and show the director orientation along the rubbing direction in the untwisted regions of the $N^*$ and $N_F^*$. The retardance increases linearly with the distance $d$ along the thickness gradient, Fig.7(c). The maximum retardance of the untwisted region in $N^*$ phase is $\Gamma_{max} = 403$ nm, which is close to the retardance of the wedge at the local thickness $h = \mathcal{P}/4 = 2.3$ μm, since $\Delta n = 0.17$ and $\mathcal{P} = 9.2$ μm. The result implies that the Moebius $\pi$-disclination that separates the untwisted and $\pi$-twisted regions is located at $h_\pi = \mathcal{P}/4$ in the $N^*$ [56]. In the $N_F^*$, the maximum retardance of the untwisted region is $\Gamma_{max} = 979$ nm, Fig.7(c), close to the retardance of wedge at $h = \mathcal{P}/2 = 5$ μm, as $\Delta n \times \mathcal{P}/2 = 990$ nm. Therefore, the first defect that separates the untwisted and $2\pi$- twisted region in the $N_F^*$, is a $2\pi$-disclination residing at $h_{2\pi} = \mathcal{P}/2$.

The $2\pi$-disclination preserves its straight shape in the magnetic field even when other defects change to zigzags, Fig.4(c). The apparent reason is that the estimated $h_{2\pi}$ is smaller than the thickness $h_t = \frac{\pi}{B}\sqrt{\frac{\mu_0 K_2}{\chi_a}}$ at which the field causes a Frederiks-like twist. If one inserts $B = B_{zz} = B_c/2$ in the last equation, then $h_t = \frac{2\mathcal{P}}{\pi}$, which is larger than $h_{2\pi} = \mathcal{P}/2$.

The polar unidirectional azimuthal anchoring at the bounding plates forces the polarization vector **P** in the $N_F^*$ to twist only by $2m\pi$, where $m$ is an integer. Any Grandjean zone with an odd number of $\pi$-twists in a parallel assembly wedge is prohibited and is removed by restructuring, which is a complex process that involves a transformation of $b = \mathcal{P}/2$ dislocations into $b = \mathcal{P}$ lines by absorption of an $N_F^*$ layer of the thickness $\mathcal{P}/2$, Fig.8.

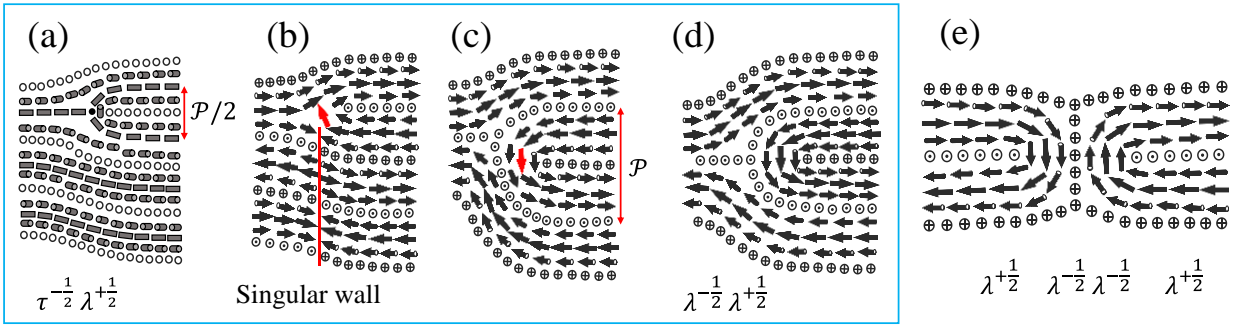

Fig.8. Nonsingular disclinations in the $N_F^*$. (a-d) Restructuring of (a) $\tau^{-\frac{1}{2}}\lambda^{+\frac{1}{2}}$ into (c,d) $\lambda^{-\frac{1}{2}}\lambda^{+\frac{1}{2}}$ disclination pair through the creation of a singular wall of polarization (b). The restructuring amounts to an addition of a half-pitch $\mathcal{P}/2$ to the Burgers vector of the dislocation, which eliminates Grandjean zones with an odd number of $\pi$-twists in a parallel assembly of a wedge; (e) Lehmann cluster $b = 0$ comprised of two $\lambda^{-\frac{1}{2}}$ and two $\lambda^{+\frac{1}{2}}$ disclinations is fully compatible with the polar ordering of the $N_F^*$.

In the N*, $\tau^{\pm\frac{1}{2}}$ disclinations are singular since the local director is perpendicular to them, while $\lambda^{\pm\frac{1}{2}}$ disclinations are nonsingular, Fig.8(a). In the $N_F^*$, the difference between these two classes is even more pronounced, as the $\tau^{\pm\frac{1}{2}}$ disclinations cannot exist as isolated defects and must be connected to a defect wall separating the opposite directions of polarization, **P** and −**P**, Fig.8(b). A $\pi$ rotation of **P** around a $\tau^{\pm\frac{1}{2}}$ disclination line yields a −**P** state incompatible with **P**. This feature leads to "composite" defects, representing $\pm 1/2$ singular disclinations connected by a wall, as predicted for the non-chiral $N_F$ [57,58] and observed in thin $N_F$ films [59]. Below the $SmZ_A^*$ - $N_F^*$ transition temperature, the emerging domain wall attached to the $\tau^{-\frac{1}{2}}$ core in Fig.8(c) must be healed, as it carries a large energy of a perturbed ferroelectric order. By restructuring the $\tau^{-\frac{1}{2}}$ core and gliding it along the z-axis, which amounts to the elimination of the singular wall and an addition of a half-pitch $\mathcal{P}/2$ to the defect, one transforms the $\tau^{-\frac{1}{2}}\lambda^{+\frac{1}{2}}$ pair into the $\lambda^{-\frac{1}{2}}\lambda^{+\frac{1}{2}}$ one with $b = \mathcal{P}$. Figure 7(a) shows that the Grandjean zones and well-defined dislocations are strongly distorted during this restructuring at the $SmZ_A^*$ - $N_F^*$ transition point.

Sometimes the networks show dislocations of a zero Burgers vector which run along the thickness gradient and connect neighboring $b = \mathcal{P}$ dislocations, Fig.2(a). At the dislocation nodes, the mechanical equilibrium requires the sum of line tensions $T$ of dislocations to be zero [2,14]. The angle $\phi_\mathcal{P}$ between the $b = 0$ line and each of the two segments of the $b = \mathcal{P}$ dislocations is 38° in the N* (100 °C) and 37° in the $N_F^*$ phase (55°). It implies that the line tension of $b = 0$ dislocation is about 1.6 times stronger than the line tension of the $b = \mathcal{P}$ defect, $T_0/T_\mathcal{P} = 2\cos\phi_\mathcal{P} \approx 1.6$ in both the $N_F^*$ and N*. The estimate is close to the one found previously, $T_0/T_\mathcal{P} \approx 1.7$, for a conventional N* [14]. The result is reasonable, since the $b = 0$ dislocation is formed by two $\lambda^{+\frac{1}{2}}$ and two $\lambda^{-\frac{1}{2}}$ nonsingular disclinations, which produce a so-called "Lehmann cluster" [14]. The existence of the Lehmann clusters supports the idea that $\lambda$ disclinations of semi-integer strength in the N* and $N_F^*$ are similar. The Lehmann cluster in Fig.8(e) is fully compatible with the ferroelectric order of the $N_F^*$.

***Dislocations upon cooling; antiparallel assembly.*** In a $N_F^*$ wedge cell with an antiparallel assembly, surface anchoring forces **P** to twist by $(2m + 1)\pi$. The untwisted region of the $N^*$ phase becomes a $\pi$-twisted region in the $N_F^*$, Fig.7(b). The $b = \mathcal{P}/2$ dislocations behave similarly to what is described above for the parallel assembly. They do not change much upon the $N^*$- $SmZ_A^*$ transition but they all transform into $b = \mathcal{P}$ in the $N_F^*$, Fig. 7(b).

***Dislocations $b = \mathcal{P}$ splitting upon heating; parallel assembly wedge.*** On heating the sample from the $N_F^*$ to the $SmZ_A^*$ at a slow rate of 0.05°C/min, $b = \mathcal{P}$ dislocations undergo a complex restructuring process at the transition point, which results in the appearance of a few $b = \mathcal{P}/2$ dislocations in the thinnest part of the $SmZ_A^*$ wedge, Fig.9. Once the heating yields the $N^*$ phase, the thick dislocations $b = \mathcal{P}$ in the thin part of the sample, $h < h_c$, split into pairs of dislocations $b = \mathcal{P}/2$. The splitting starts at the thinner part of the wedge.

As already stated, the cores of the $b = \mathcal{P}$ and $b = \mathcal{P}/2$ dislocations are very different, the first being non-singular and the second containing a singular $\tau$ disclination. The core energy $E_{c,\tau\lambda}$ of a $b = \mathcal{P}/2$ dislocation is estimated as [14]: $E_{c,\tau\lambda} \approx \frac{\pi}{2} K ln \frac{\mathcal{P}}{4r_c}$, where $r_c$ is the radius of the $\tau$ core, on the order of 1-10 molecular sizes. For typical $r_c$=10 nm and $\mathcal{P} = 10$ μm, $E_{c,\tau\lambda}$ could be as large as $\sim 17K$. In contrast, the $b = \mathcal{P}$ core is smooth over the area $\sim \mathcal{P}^2$ and the core energy is $E_{c,\lambda\lambda} \approx K$ [14]. At first sight, the strong inequality $E_{c,\tau\lambda} \gg E_{c,\lambda\lambda}$ prohibits splitting of $b = \mathcal{P}$ dislocations into two $b = \mathcal{P}/2$ dislocations. The paradox is resolved by noticing that the core energies should be supplemented with the compression energy of $N^*$ pseudolayers. Briefly, inserting a slab of a thickness $b = \mathcal{P}/2$ into a cell of a fixed thickness $h$ requires less compression energy as compared to a thicker $b = \mathcal{P}$ slab. The difference is significant only when the number $n = 2h/\mathcal{P}$ of pseudolayers of thickness $\mathcal{P}/2$ each is small. As demonstrated in Ref. [14], the difference between the line energies of two $b = \mathcal{P}/2$ and one $b = \mathcal{P}$ dislocations, which accounts for the compressibility term, writes

$$\frac{(2E_{\mathcal{P}/2} - E_\mathcal{P})}{K} \approx \pi ln \frac{\mathcal{P}}{4r_c} - \frac{\mathcal{P}^3}{32h\xi^2 \tan\alpha}, \tag{4}$$

where $\xi = \frac{\mathcal{P}}{2\pi}\sqrt{\frac{3K_{33}}{8K_{22}}}$ is the "penetration" length, which is usually a few times smaller than the pitch $\mathcal{P}$ [60]. This relationship explains why in the thin part of the wedge, the $b = \mathcal{P}/2$ dislocations are more stable than the $b = \mathcal{P}$ defects once the material transitions into the N*. The energy gain increases for smaller $h/\mathcal{P}$, in agreement with the experimental observation that the splitting occurs first in the thinnest part of the wedge, Fig.9. Furthermore, the critical thickness can be estimated from the condition $E_{\mathcal{P}/2} \approx E_{\mathcal{P}}$ as $h_c \approx \frac{\mathcal{P}^3}{32\alpha\xi^2\pi\ln\frac{\mathcal{P}}{4r_c}}$. With the known $\alpha = 1.24°$, $\mathcal{P} = 10$ μm for M0.2 and the estimates $r_c=10$ nm, $\xi = 0.2\mathcal{P}$, one obtains $h_c = 21$ μm, in a remarkable agreement with the experimental value above.

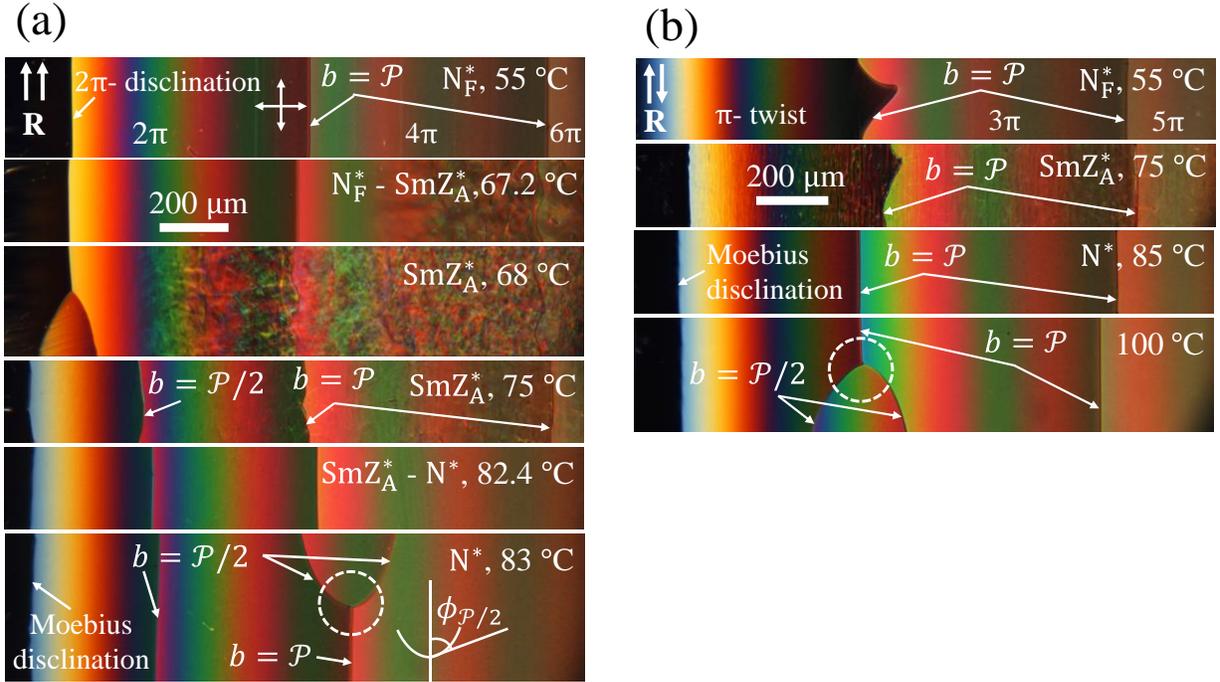

Fig.9. Edge dislocations in the wedge heated from the $N_F^*$ to $SmZ_A^*$ and N*; (a) parallel assembly wedge, $\alpha = 0.80°$; (b) antiparallel assembly, $\alpha = 0.77°$. The thick $b = \mathcal{P}$ dislocations restructure and split into two thin edge dislocations $b = \mathcal{P}/2$. Mixture M0.2.

### E. Twisted planar alignment of chiral $SmZ_A^*$.

In the non-chiral $SmZ_A$ phase, the polarization **P** is parallel to the smectic layers, reversing its polarity from one layer to the next. In confinement, an achiral $SmZ_A$ can adopt two different alignment geometries [32]. In the chiral $SmZ_A^*$, the orientation of smectic layers is better defined: they should be perpendicular to the helicoidal axis $\hat{\chi}$ to preserve equidistance. To verify this statement, we study the response of the $\pi$-twisted N$^*$ and SmZ$_A^*$ to an in-plane electric field applied perpendicularly to $\hat{\chi}$ and along the buffing direction, Fig. 10.

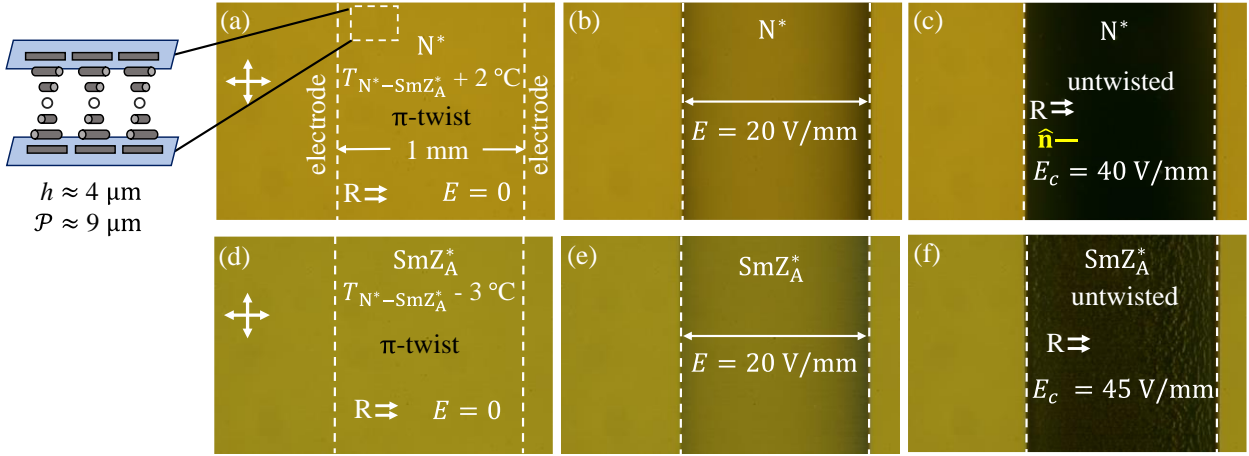

Fig. 10. Unwinding of a $\pi$-twisted (a-c) N$^*$, (d-f) SmZ$_A^*$ in planar cells with parallel assembly by an in-plane electric field applied along the buffing direction. Mixture M0.2, $h = 4.1$μm.

The $\pi$-twisted N$^*$ structure undergoes helix unwinding, Fig. 10(a-c), at some critical field $E_c = 40$ V/mm. Similarly, the $\pi$-twisted SmZ$_A^*$ unwinds at $E_c = 45$ V/mm, Fig. 10(d-f), which means that the smectic layers are perpendicular to $\hat{\chi}$. This verifies that the twisted Grandjean zones in the wedge SmZ$_A^*$ cells have a twisted planar alignment.

## IV. CONCLUSION

Below, we formulate the main results in relationship to the three explored phases, N$^*$, SmZ$_A^*$, and N$_F^*$.

**Paraelectric cholesteric N$^*$**. The Grandjean-Cano structures in the cholesteric N$^*$ phase of the ferroelectric material DIO are the same as previously established in the optical [35] and FCPM [14] experiments on conventional N$^*$ materials. Namely, the thickness gradient is relieved by a one-dimensional lattice of dislocations of Burgers vector $b = \mathcal{P}/2$ in the thin part of the wedge, $h < h_c$, and $b = \mathcal{P}$ in the thick part, $h > h_c$.

The separation between the dislocations allows us to determine the temperature dependence of the helical pitch $\mathcal{P}$, which shows a weakly non-monotonous behavior upon cooling, with a minimum in the middle of the N* temperature range. $\mathcal{P}$ remains finite during the N*-SmZ$_A^*$ transition. The N* pitch diverges in a magnetic field applied in the plane of the wedge, perpendicularly to the dislocations. The full unwinding is achieved at about $B_c \approx 1$ T.

The dislocations in the N* phase show cores split into pairs of disclinations, as foreseen by the Kleman-Friedel model [8,9]. Namely, a $b = \mathcal{P}/2$ core splits into a pair of a singular $\tau^{-\frac{1}{2}}$ and a nonsingular $\lambda^{+\frac{1}{2}}$ disclinations separated by a distance $\approx \mathcal{P}/4$, while a $b = \mathcal{P}$ core splits into a nonsingular $\lambda^{-\frac{1}{2}}$ and $\lambda^{+\frac{1}{2}}$ separated by $\mathcal{P}/2$. An in-plane magnetic field $\sim B_c/2$ causes zigzag instability of $\lambda^{-\frac{1}{2}}\lambda^{+\frac{1}{2}}$ cores, but the $\tau^{-\frac{1}{2}}\lambda^{+\frac{1}{2}}$ cores remain rectilinear. Because of the singular structure, the core energy of the $b = \mathcal{P}/2$ dislocations is much higher than that of the nonsingular $b = \mathcal{P}$ dislocations, which makes the existence of $b = \mathcal{P}/2$ dislocations at $h < h_c$ puzzling. Their stability is explained by the finite compressibility of cholesteric pseudolayers: in the thin part of the wedge, the compressibility energy cost of inserting a slab of a thickness $b = \mathcal{P}$ is much higher than the compressibility cost of inserting a slab of a thickness $b = \mathcal{P}/2$. The balance of curvature and compressibility energies defines $h_c$ [14]. Note that the existence of two different types of dislocations must be accounted for in the measurements of $\mathcal{P}$ in Grandjean-Cano wedges.

**Antiferroelectric chiral SmZ$_A^*$.** Upon cooling from the N*, the SmZ$_A^*$ preserves the dislocations with the Burgers vector $b = \mathcal{P}/2$ in the thin and $b = \mathcal{P}$ in the thick part of the sample. They also split into the $\tau^{-\frac{1}{2}}\lambda^{+\frac{1}{2}}$ and $\lambda^{-\frac{1}{2}}\lambda^{+\frac{1}{2}}$ pairs, respectively. On heating from the N$_F^*$, $b = \mathcal{P}$ dislocations split into two $b = \mathcal{P}/2$ lines in the thin part of the wedge, $h < h_c$. The SmZ$_A^*$ dislocations show no zigzag instability in the magnetic field as high as 1 T. Furthermore, the helical pitch remains unchanged in the same range of the field. We associate these features with the layered structure of the SmZ$_A^*$,

**Ferroelectric cholesteric N$_F^*$.** Cooling from SmZ$_A^*$ into the N$_F^*$ causes a dramatic restructuring, in which dislocations $b = \mathcal{P}/2$ transform into $b = \mathcal{P}$ dislocations. The $b = \mathcal{P}$ dislocations are similar to their counterparts in the N*. The $b = \mathcal{P}$ dislocations in all explored phases carry a split nonsingular $\lambda^{-\frac{1}{2}}\lambda^{+\frac{1}{2}}$ core, which is compatible with the polar ordering of N$_F^*$, antiferroelectric ordering of $\text{SmZ}_\text{A}^*$, and apolar ordering of $\text{N}^*$. Further evidence of the stability of $\lambda$ disclination is the existence of $b=0$ dislocations; their cores contain two $\lambda^{-\frac{1}{2}}$ and two $\lambda^{+\frac{1}{2}}$ disclinations (the so-called "Lehmann clusters"). An interesting feature of the antiparallel assembly $\text{N}_\text{F}^*$ wedge is that the first line defect in the thinnest part is of a wavy shape even when there is no magnetic field; this shape is related to the splay-cancellation effect. The bound charge caused by the confinement-induced splay of polarization can be reduced by an additional splay in the cell plane.

The transformation of thin dislocations $b=\mathcal{P}/2$ inherited from the antiferroelectric $\text{SmZ}_\text{A}^*$ into thick $b=\mathcal{P}$ dislocations in the $\text{N}_\text{F}^*$ is necessitated by the fact that the $\tau^{\pm\frac{1}{2}}$ disclinations, unlike their $\lambda^{\pm\frac{1}{2}}$ counterparts, cannot exist as isolated defects and must be connected to a domain wall in the polarization field. These walls, however, can be eliminated by the reconstruction of the defect core, which is equivalent to the addition of a half-pitch $\mathcal{P}/2$ to the $\tau^{-\frac{1}{2}}\lambda^{+\frac{1}{2}}$ pair and its transformation into $\lambda^{-\frac{1}{2}}\lambda^{+\frac{1}{2}}$, with a corresponding doubling of the Burgers vector.

The experiments demonstrate that the structure and field response of the confinement-induced edge dislocations depend strongly on the type of microscopic ordering, which changes from paraelectric to antiferroelectric and ferroelectric. This dependency can be used not only in measuring the properties such as the helical pitch, but also in identifying the type of ordering in newly synthesized materials.

## Acknowledgments

We thank Sergij V. Shiyanovskii for useful discussions. The work was supported by NSF Grant ECCS-2122399 (ODL) and NSF Grant DMR-2210083 (JTG).